\documentclass[preprint]{aastex}
\usepackage{epsfig}

\shorttitle{GAUDI: a preparatory archive for the COROT mission}
\shortauthors{Solano et al.}

\begin{document}

\title{GAUDI: a preparatory archive for the COROT mission \footnote{Based on observations collected at La Silla (ESO proposals 67.D-0169, 69.D-0166 and 70.D-0110), Telescopio Nazionale Galileo (proposal 6-20-068), Observatoire de Haute-Provence, South African Astronomical Observatory, Tautenburg Observatory and Sierra Nevada Observatory.}}

\author{E. Solano\altaffilmark{1}, C. Catala\altaffilmark{2}, R. Garrido\altaffilmark{3}, E. Poretti\altaffilmark{4}, E. Janot-Pacheco\altaffilmark{5}, R. Guti\'{e}rrez\altaffilmark{6}, R. Gonz\'{a}lez\altaffilmark{7}, L. Mantegazza\altaffilmark{4}, C. Neiner\altaffilmark{8,9}, Y. Fremat\altaffilmark{9}, S. Charpinet\altaffilmark{10}, W. Weiss\altaffilmark{11}, P. J. Amado\altaffilmark{3}, M. Rainer\altaffilmark{4}, V. Tsymbal\altaffilmark{12}, D. Lyashko\altaffilmark{12}, D. Ballereau\altaffilmark{9}, J.C. Bouret\altaffilmark{13}, T. Hua\altaffilmark{13}, D. Katz\altaffilmark{9}, F. Ligni\`eres\altaffilmark{10}, T. L\"uftinger\altaffilmark{11}, P. Mittermayer\altaffilmark{11}, N. Nesvacil\altaffilmark{11}, C. Soubiran\altaffilmark{14}, C. Van't Veer-Menneret\altaffilmark{9}, M.J. Goupil\altaffilmark{2}, V. Costa\altaffilmark{3}, A. Rolland\altaffilmark{3}, E. Antonello\altaffilmark{4}, M. Bossi\altaffilmark{4}, A. Buzzoni\altaffilmark{15}, C. Rodrigo\altaffilmark{1}, C. Aerts\altaffilmark{16}, C.J. Butler\altaffilmark{17}, E. Guenther\altaffilmark{18}, A. Hatzes\altaffilmark{18}} 

\altaffiltext{1}{INSA-LAEFF, VILSPA, P.O. Box 50727, 28080 Madrid, Spain}
\altaffiltext{2}{LESIA / UMR 8109 du CNRS, Observatoire de Paris, France}
\altaffiltext{3}{Instituto de Astrof\'{\i}sica de Andaluc\'{\i}a, C.S.I.C., Granada, Spain}
\altaffiltext{4}{INAF-Osservatorio Astronomico di Brera, Merate, Italy}
\altaffiltext{5}{Instituto de Astronomia, Geofisica e Ciencias Atmosfericas da Universidade de Sao Paulo, Brasil}
\altaffiltext{6}{INTA-LAEFF, VILSPA, P.O. Box 50727, 28080 Madrid, Spain}
\altaffiltext{7}{XMM-Newton Science Operation Centre, ESA, VILSPA, P.O. Box 50727, 28080 Madrid, Spain}
\altaffiltext{8}{RSSD, ESTEC / ESA, Noordwijk, The Netherlands}
\altaffiltext{9}{GEPI/UMR 8111 du CNRS, Observatoire de Paris, France}
\altaffiltext{10}{Laboratoire d'Astrophysique de Toulouse et Tarbes, UMR 5572 du CNRS, France}
\altaffiltext{11}{Institut f\"ur Astronomie, Universit\"at Wien, Austria}

\altaffiltext{12}{Tavrian National University, Simferopol, Ukraine}

\altaffiltext{13}{Observatoire Astronomique Marseille-Provence, Marseille, France}
\altaffiltext{14}{Observatoire de Bordeaux, France}
\altaffiltext{15}{INAF-Osservatorio Astronomico di Bologna, Italy}
\altaffiltext{16}{Instituut voor Sterrenkunde, Katholieke Universiteit Leuven, Leuven, Belgium}
\altaffiltext{17}{Armagh Observatory, United Kingdom}
\altaffiltext{18}{Th\"uringer Landessternwarte Tautenburg, Germany}

\begin{abstract}

The GAUDI\footnote{http://sdc.laeff.esa.es/gaudi/} database (Ground-based
Asteroseismology Uniform Database Interface) is a preparatory archive for the
COROT\footnote{http://www.astrsp-mrs.fr/projets/corot/} (COnvection,  ROtation
and planetary Transits) mission developed at
LAEFF\footnote{http://www.laeff.esa.es} (Laboratory for Space Astrophysics and
Theoretical Physics). Its intention is to make the ground-based observations obtained in
preparation of the asteroseismology programme available in  a simple and
efficient way. It contains spectroscopic and photometric data together 
with inferred physical parameters for more than 1\,500 objects gathered since January 1998 in 6 years of 
observational campaigns. In this paper, the main functionalities and
characteristics of the  system are described.

\end{abstract}


\section{Introduction}

The COROT satellite \citep{baglin} will be launched in 2006 and is intended to perform high-precision (micromagnitude) photometric monitoring of stellar targets to achieve two
main objectives: 

\begin{itemize}
\item Asteroseismology of about a hundred dwarf stars to give direct information
on the structure and dynamics of their interiors; among those, a few bright stars 
(F,G dwarfs, $\beta$ Cep, and $\delta$ Sct; V $\leq$ 6.5) will be monitored for up to 5 months, and up to 10 fainter (V $\leq 9.5$) stars per field will be observed simultaneously, so as to cover the HR diagram
as completely as possible. 

\item The detection of planets down to Earth-sized telluric planets, using 
the method of transits. 
\end{itemize}
In addition to this, the mission will provide accurate, continuous, 
photometric monitoring of thousands of fainter stars (11.5 $\leq$ V $\leq$ 16). 
Such space-based photometry will have a signal--to--noise ratio several orders of magnitude 
better than can be obtained using ground-based facilities and will provide a mass of 
highly original data on a wide diversity of stars.

The intrinsic nature of the seismology programme of the COROT mission (very
long observations of a small number of bright stars) makes  the target
selection a critical issue. In order to take full advantage of the COROT data
and to strongly constrain stellar evolution models, the seismic information
needs to be complemented with precise and reliable knowledge of  the
fundamental stellar physical parameters (i.e. effective temperatures,
luminosities, surface  gravities, rotational velocities and chemical
abundances). Also, it is necessary to identify potential doubled-lined
spectroscopic binaries (SB2), due to the difficulty of disentangling the 
individual oscillation information from the composite spectra. Finally,
peculiarities such as photometric or spectroscopic variability, magnetic
activity, spectral line asymmetries, etc... also need to be identified. For an
optimal definition of the final target list  it is, therefore, essential to
gather as much a priori information as possible on the physical parameters and
characteristics of the stars. However, it was soon realized that the available
information for many of the potential  targets was insufficient for a reliable
selection. For this reason an ambitious ground-based  observing
programme, to obtain Str\"{o}mgren photometry and high resolution 
spectroscopy in order to accurately and reliably determine the physical
parameters for more than 1\,500 objects, was launched within the framework of
the COROT mission.

To cope with this vast and heterogeneous (different instrumentation, reduction procedures,  analysis techniques, people,...) dataset in a convenient way it
was necessary to develop a user-friendly access system. This was considered as
a fundamental objective by all the  scientific groups involved in the COROT
project and, in March 2001, LAEFF was nominated as responsible  for the
development and long-term maintenance of GAUDI.

\section{The data}

The requirement for long and continuous observations on the same field 
imposes the necessity for a polar orbit for the satellite, with the line
of sight almost perpendicular to the orbital  plane to avoid eclipses from the
Earth. As the line of sight has to be in opposition  with the Sun, every six
months the satellite will be rotated by 180\degr. In April 2001,  the COROT
scientific council made the final decision for the orbital plane of the 
satellite defining an accessibility zone centered at $\alpha$=6h50m and 18h50m
and $\delta$=0\degr, and about 10\degr $ $ in radius. This zone was named scenario 4 to distinguish it from other observing windows previously considered. In a first stage the catalogue
contained all stars at a distance less than  10 degrees from the centre of the
accessibility zone with a visual apparent magnitude brighter  than V=8.
Whenever possible, giants were excluded on the basis of their Hipparcos 
parallaxes. It was then noticed that limiting the input catalogue to objects
brighter than V=8 was sometimes insufficient to take full advantage of 
the instruments capabilities (to monitor up to 10 targets simultaneously in the
seismology field) as there  were not enough faint candidates around the bright
main targets. To overcome this situation an extension of the spectroscopic and
photometric campaigns was recently initiated to cope with the stars with 8
$\leq$  V $\leq$ 9.5 in the fields of the COROT main targets. All the new
observational data as well  as the physical parameters obtained in the
characterization study will be included in the  GAUDI archive in the
coming months leading to an increase of about 50$\%$ in the present 
content of the database. 

\subsection{Echelle Spectroscopy}  Most of the spectroscopic observations were
conducted on telescopes of the 2m and 4m class, equipped with high  resolution
echelle  spectrographs (R = 40\,000 -- 50\,000) from three different sites (OHP
with the ELODIE  instrument on the 1.93m telescope, La Silla with FEROS on the
1.52m and 2.2m telescopes and La Palma  with the SARG spectrograph on the 3.5m
TNG). Typical signal-to--noise ratios range  from 100 to 150 at 5500 \AA. The
wavelength interval covered was 3\,900--6\,800 \AA, 3\,800--9\,100 \AA $ $ and
4\,600--6\,800 \AA $ $ for ELODIE, FEROS and SARG respectively. 

In addition to these three main sites, a few more spectra were secured at La
Silla with the 1.2m Swiss telescope equipped with the CORALIE spectrograph, at
SAAO (South Africa) with the 1.9m Radcliffe telescope equipped with the GIRAFFE
echelle spectrograph, and at Tautenburg (Germany) with the 2m telescope and
Coud\'e spectrograph.

For the reduction of spectroscopic data acquired with the ELODIE, FEROS and
CORALIE spectrographs, the first steps (order localization, background estimate
and subtraction and wavelength calibration) were performed using available on-line
reduction pipelines. Special attention was paid to the blaze and flat-field
correction. Instead of the standard correction implemented in the available
pipelines, we used the following procedure: (1) one or several spectra of 
O-type stars were acquired during the observations. (2) The extracted orders of
these reference spectra were used to define the blaze function by fitting
low-order cubic splines to sets of data points across each order, carefully
avoiding regions containing the few spectral lines in these spectra. The
extracted spectra of our programme stars were corrected separately for
pixel-to-pixel response using Tungsten flat-field spectra and from the blaze
response as described above. Such procedures led to significant improvements on the goodness of the results, also in those cases with well settled reduction 
pipelines (for instance, see Rainer (2003) for FEROS). 

Special care was taken with the flat-field correction of SARG data given the different sensitivity of the spectral output depending on the position angle of the image derotator. Furthermore, periodic and spurious signals due to CCD electronics and affecting some of the spectra, prevented a standard cleaning procedure via FF techniques, requiring instead an optimized fitering and data extraction (see Tsymbal et al. (2003) for full details on the reduction procedure).

The data from Tautenburg were reduced using the IRAF package with a standard spectroscopic reduction pipeline whereas for the SAAO GIRAFFE data, we used the Esprit reduction package as described in Donati et al. \cite{donati}. GIRAFFE is a copy of the MUSICOS spectrograph, fully described in Baudrand \& Bohm \cite{Baudrand}.

In addition to the reduced echelle spectra, the mean photospheric line
profiles  of the stars were computed following the LSD (Least Square
Deconvolution) method described in Donati et al. \cite{donati}. In this method,
a line pattern function is constructed, containing all the lines supposedly
present in the spectrum as Dirac functions, with heights set to the central
line depths as calculated by the SYNTHE programme \cite{kur}. The observed
spectrum is then deconvolved with this line pattern function, yielding a "mean"
photospheric line profile. This method has proved to be a powerful tool
to calculate accurate rotational velocities as well  as to detect multiple
systems, line asymmetries and spectroscopic anomalies.

All reduced spectra and mean profile files were recorded as standard FITS binary tables, with a normalized header including all necessary information on the object, on the instrument, on the exposure, and on the reduction. For FEROS and ELODIE spectra the binary table contains five columns with information on wavelength, flux (both un-normalized and normalized), signal--to--noise ratio and echelle order. The SARG spectra also contains the same five columns but with null values in the un-normalized flux and the echelle order columns.

\subsection{Str\"omgren photometry}

Observations on the $ubvy\beta$ system were obtained over two-week 
runs for each summer and winter observing periods during the years
    from 2000 to 2004.  The fully-automated six-channel ($uvby\beta$)
    spectro-photometer on the 0.9-m telescope at the Sierra Nevada
    Observatory (OSN) was used for these observations.  In the paper by P.J. Amado (in preparation)  details of the observing, reduction
    and transformation procedures are given in full.  The telescope, the
    photometer, the autocentring process and the data acquisition are
    all taken care of by {\sc telestrom}, a software package developed
    at the Instituto de Astrof\'{\i}sica de Andaluc\'{\i}a.

    Observations of programme stars were taken interleaved with those of standard
    stars.  Between three and six standards were observed once every
    one or one and a half hours during the night to follow and
    determine the extinction.  Sky background observations were made
    according to the Moon's phase and its position in the sky.  On nights
    with no or very low sky background flux, one sky observation was
    taken with the extinction stars. This number was increased for nights
    with higher sky flux, with up to two sky measurements per star, one before
    and one after the measurement of the star.

    Transformation to the standard system followed Gronbech et al.
    \cite{Gron76} with the standard system defined by stars
    selected from the catalogues of Olsen \cite{Olse93,Olse94V,Olse94}.

\subsection{Data policy}

Data in the GAUDI archive become public to the world community after one year of proprietary time starting in January
2003 for data delivered before this date and at the time of ingestion into GAUDI if this 
happened after January 2003. This means that the first release of public data took place in 
January 2004. At the time of writing, 851 spectra of 433 objects out of a total of 2\,369 spectra as well as Str\"omgren photometry for 1\,407 objects are publicly available.

Public data are free with no restrictions. Any user connected to the internet
is able to  query the archive. Private data are available only for people
involved in the preparation of the COROT mission. They must be registered and
identify  themselves when logging into the system using a username and password
provided by the GAUDI staff. Information on the release dates of the private
datasets can be obtained from the welcome page of the GAUDI system. Research work benefitting from the use of both public and private GAUDI data should include the following acknowledgment in publications: ''Based on
GAUDI, the data archive and access system of the ground-based asteroseismology
programme of the COROT mission. The GAUDI system is maintained at LAEFF which 
is part of the Space Science Division of INTA''.

\section{The GAUDI system. Functionalities}

Data archiving typically comprises the data ingestion, data storage, 
data management and data retrieval through an interface. Once the data were reduced they were shipped to LAEFF for their ingestion in the GAUDI archive. The reduced spectroscopic data were adapted to the spectral data model defined for GAUDI. To ensure 
integrity, metadata were extracted from the FITS headers of the spectroscopic data and 
ingested in the GAUDI database in an automated way. The data themselves reside on a mass storage
system (magnetic disks) in FITS format. Data safekeeping is guaranteed with a well-defined 
backup policy. Moreover, a number of quality control tests have been defined to ensure the 
reliability of the data and metadata provided by the archive. 

The friendliness of the user interface is extremely important for the archive
to be  effectively used. With this aim, GAUDI is HTML-based to allow straight
forward access (no need to implement special software on the user's side) through the Web.

\subsection{Archive search}

\begin{figure}[h]
\begin{center}
\epsfig{file=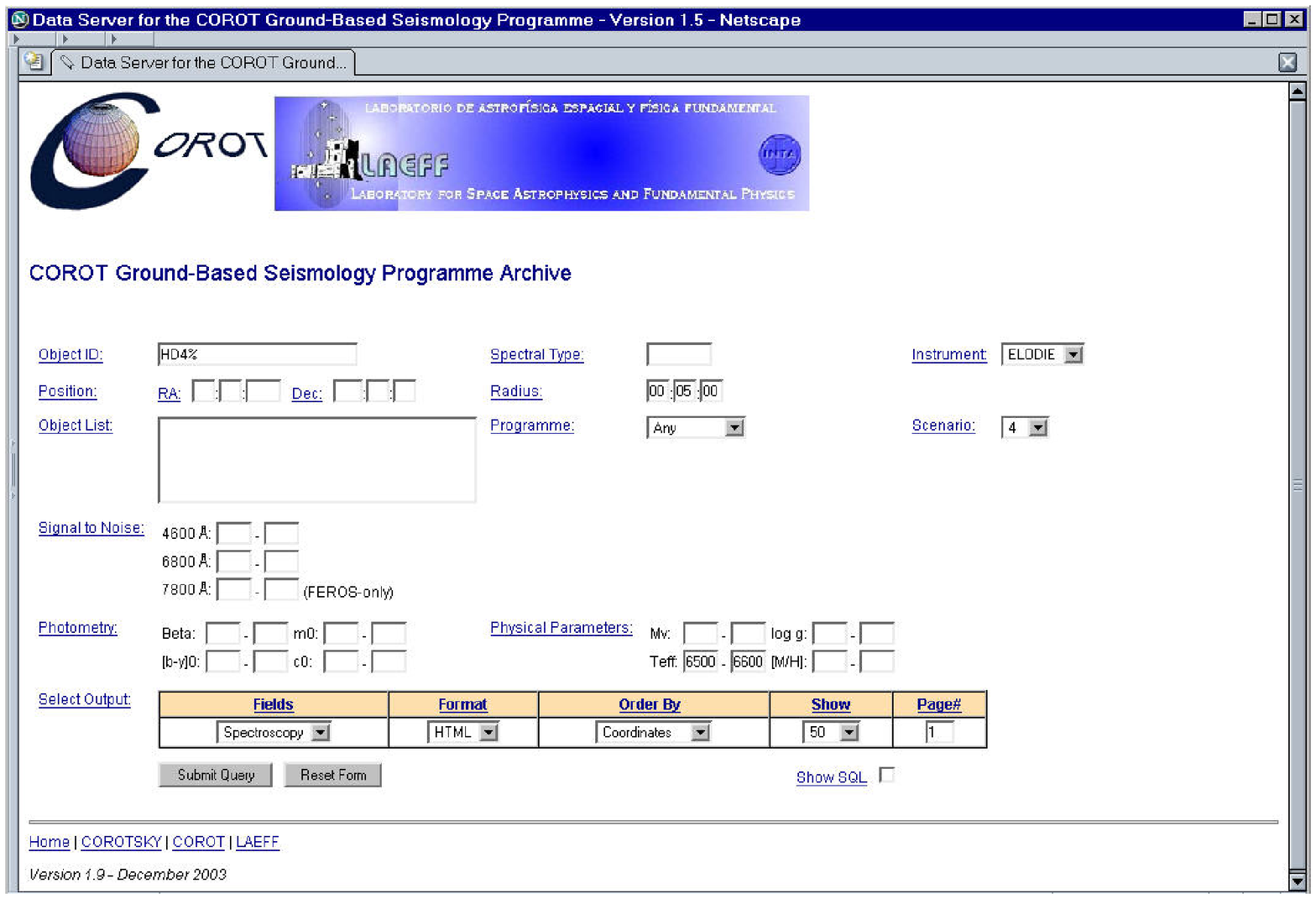, width=15cm}
\end{center}
\caption{GAUDI search capabilities (See Sect. 3.1 for details). \label{search}}
\end{figure}

 The query to  the access catalogue is made by means of a fill-in form. In
addition to the  "classical" query keywords (object name or list of names and
coordinates), GAUDI  allows project-related interrogation of the system
(observing scenario, instrument, programme) as well as the  possibility to
explore the spectroscopy (signal--to--noise ratio), photometry (dereddened
color  indexes) and stellar physical parameter space (spectral type,
effective temperature, surface gravity,  absolute magnitudes, metallicity)
(Fig.~\ref{search}). Searches are case independent and wildcards  are
permitted. The system also incorporates a built-in name  resolver allowing
queries by any of the names provided by the SIMBAD database.

The three output fields (spectroscopy, photometry and physical parameters) can
be presented  in different formats (HTML, ASCII, tab-separated or
comma-separated values) and ordered by different criteria (coordinates, object,
spectral type, programme and scenario). As stated in Section 2.3, only the
spectroscopic and photometric data (and not the physical parameters) are
presently accessible from the public interface. 

\subsection{Result from search}

The information available in GAUDI is divided into three categories: spectroscopy, 
photometry and physical parameters. 

\subsubsection{Spectroscopic field} 

\begin{figure}[h]
  \begin{center}
    \epsfig{file=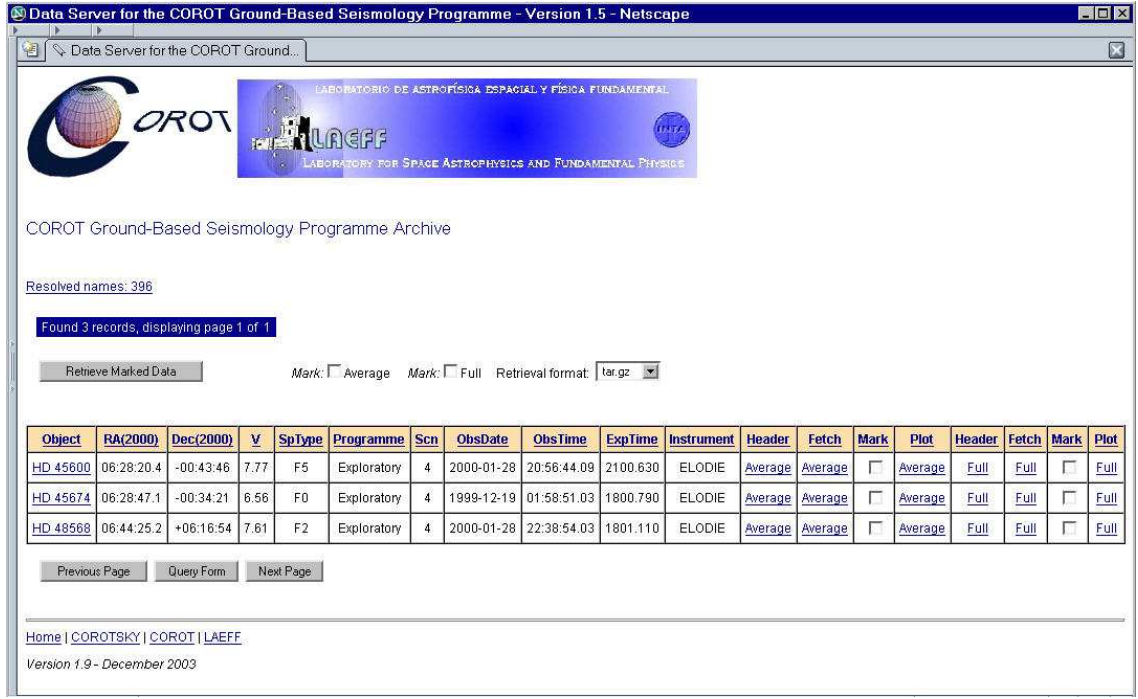, width=15cm}
  \end{center}
\caption{Result of the search displayed in Fig.~\ref{search}. For this example, the spectroscopic output field in HTML format was chosen.\label{result}} 
\end{figure}

For each observation, the spectroscopy field provides
information on the 
object name, the coordinates, the visual magnitude and spectral type, programme, 
scenario, observing date and time, exposure time and the instrument used for the observation. 
In addition to this, the following utilities are provided if the HTML output format is 
selected  (Fig.~\ref{result}).

\begin{figure*}[h]
  \begin{center}
    \epsfig{file=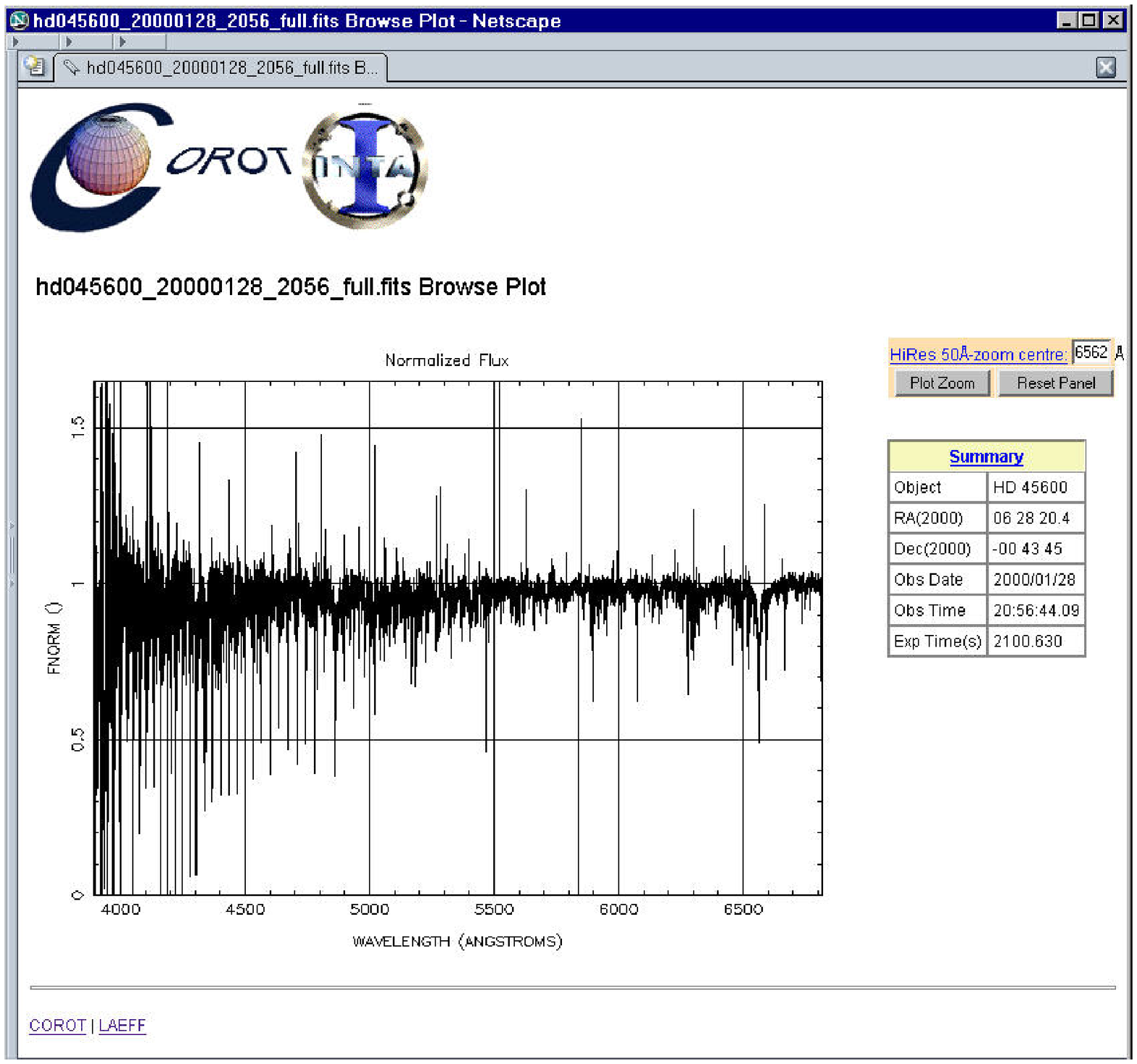, width=7cm}
    \epsfig{file=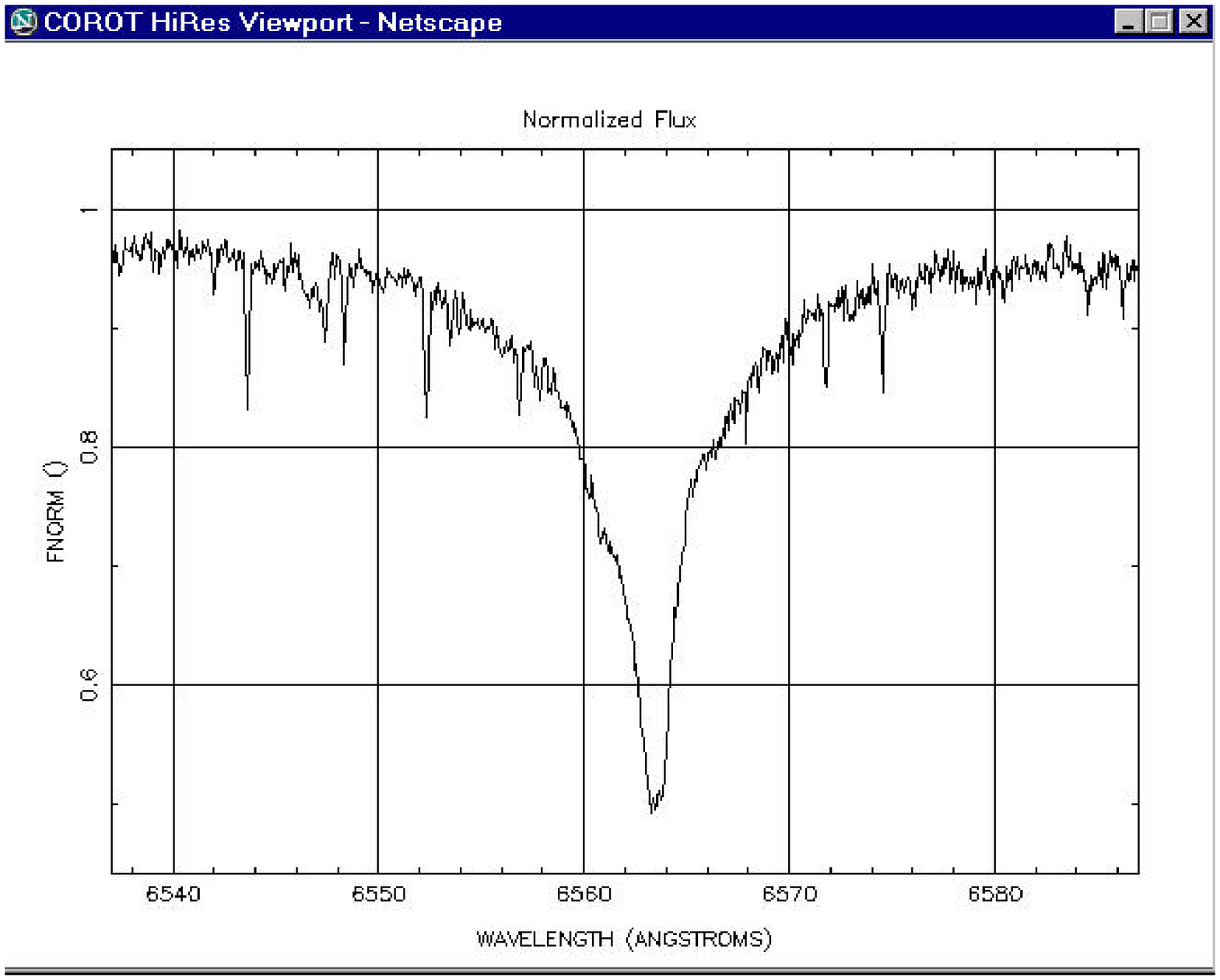, width=6cm}
  \end{center}
\caption{GAUDI data previewing capabilities. A high resolution spectrum is shown on the left. The shaded box allows a detailed view of part of the data (here the H$\alpha$ region) which is shown on the right.\label{plot}} 
\end{figure*}

\begin{itemize}
\item Link to SIMBAD: by clicking the object name, the information contained in the SIMBAD 
database is displayed.
 
\item Spectral retrieval: Spectra may be retrieved individually or in groups. For multiple 
retrieval it is possible to include/exclude individual spectra. Multiple spectra retrieval 
generates a file in either zip or tar format that can be compressed for network efficiency. 
Single spectra are retrieved uncompressed.  

\item FITS Header Display: Links are provided to display the FITS primary and binary 
table headers of each requested echelle spectrum or mean photospheric line profile file. 

\item Data previews: A browse plot of the echelle spectrum as well as the associated mean photospheric line profile is generated by clicking on the corresponding link. A panel 
summarizing the observation is displayed next to the plot, and the full FITS header can be 
listed from there. Zoom plots of 30 \AA $ $ or 30 kms$^{-1}$ (depending on whether an echelle or a mean photospheric line profile file is displayed), may be generated by entering the desired central wavelength or
radial velocity displacement. A new viewport is created allowing an overview of the entire 
set of data and a simultaneous view of the selected region. This viewport is automatically 
refreshed for subsequent zooms. A copy of a browse or zoom plot can be saved as a GIF file  (Fig.~\ref{plot}). 
\end{itemize}

\subsubsection{Photometric field}  For each observation, the photometric field
provides information  on the coordinates, visual magnitude, the (B-V) color,
programme, scenario, observing  date and time, the airmass, the Str\"omgren
indices $\mathrm{m}_{1}$, $\mathrm{c}_{1}$, $\beta$, (b-y) and the  dereddened
values $\mathrm{(b-y)}_{0}$, $\mathrm{m}_{0}$, $\mathrm{c}_{0}$,
$\mathrm{dm}_{0}$, $\mathrm{dc}_{0}$ and their corresponding errors. Dereddened indices have been obtained
using the TEMPLOGG package \citep{rogers}.

\subsubsection{Physical parameters}

For a given object, the physical parameter field gives information on
coordinates, visual magnitude, the (B-V) color, programme, scenario, spectral
type, luminosity class, proper motion, radial velocity, projected rotational
velocity, absolute magnitude, effective temperature, surface gravity and
metallicity. The physical parameters have been obtained using different
methods. GAUDI provides information on the "best" value as well as an error
estimate, the method used  and eventual comments associated to the
measurement. The best value was defined using a hierarchical scheme agreed
within the COROT project. Information on the adopted hierarchy can be obtained
by clicking on the corresponding column label.

\subsection{On-line documentation and HelpDesk} For the system to be properly
used, it must include well-structured, and have easy-to-find,   documentation both on the
COROT project and the GAUDI system. In order to efficiently answer the user's
needs, the following multi-layer approach (from the most general to the most
specific  questions) has been adopted:

\begin{itemize}
\item On-line access to project documentation: a detailed description of the project, the 
archive and the access system is given on-line from the GAUDI welcome page. Links to the COROT project and LAEFF web pages
are also available. 

\item On-line help: help on a specific keyword of the system query form can be obtained by 
simply clicking on it. 

\item Helpdesk: for those questions not channeled through the previous levels
and to  provide a continuous support to the archive users.

\end{itemize}

\section{GAUDI in the framework of the Virtual Observatory}

Although astronomical archives constitute a basic tool for modern Astrophysics
as revealed by their intensive usage, the efficiency in information
retrieval is seriously limited by the lack of interoperability among them. The
Virtual Observatory\footnote{http://www.ivoa.net} (VO) is an international
project aimed at solving the problems that this lack of interoperability
creates for multiwavelength astronomy. Accordingly, one of the VO main
objectives is the creation of a federation of astronomical Data Centres that,
with the implementation of new technologies and standards, provides an easy and
efficient access to the astronomical data. GAUDI is part of the Spanish Virtual
Observatory\footnote{http://svo.laeff.esa.es} and, as such, has been designed
following the standards and requirements defined in the framework of the
Virtual Observatory. This will permit a transparent access to the archives and
databases that will form the EURO-VO Data Centre Alliance (DCA), a
collaborative and operational network of European data centres which, by the
uptake of new VO technologies and standards, will publish data, metadata and
services in a VO-compliant way. 

One of the VO requirements that GAUDI already incorporates is the SSA (Simple Spectral Access) protocol, a standard defined for retrieving spectroscopic data from a repository of astronomical data. Through 
this method, a client searches for available data that match certain client-specified 
criteria using a HTTP GET request. The response is a table (in VOTable\footnote{http://cdsweb.u-strasbg.fr/doc/VOTable} format, an XML format defined for the exchange of tabular data in the context of the Virtual Observatory) describing 
the available data including metadata and access references (implemented as URLs) for 
retrieving them. 

\section{Conclusions}

The characterization of the COROT fields requires a lot of resources. A
well-designed, properly-implemented data archive and access system like GAUDI is demonstratively a major contribution towards the full exploitation of
these expensive-to-obtain observational data. Homogeneity and uniformity have
been two basic requirements for GAUDI. This makes it possible to conduct global archive
searches in order to confirm/discard characteristics associated to a given
group of objects. A good example of this is the discovery of 17 new Be stars
\citep{neiner} based on GAUDI data. Moreover, the advent of the Virtual
Observatory, an initiative to allow global electronic access to available
astronomical data, both space-- and ground--based, will boost the use of
astronomical archives. GAUDI, designed to fulfill the VO requirements, will
access a massive amount of different datasets (images, spectra, catalogues)
covering the sky at all wavelengths which will allow very efficient real multiwavelength research. 

The contents of GAUDI will increase in the near future by the ingestion of new photometric and spectroscopic data as well as some photometric monitoring 
data described in Poretti et al. \cite{Poretti}.
\acknowledgments

The development of GAUDI has been supported by the Spanish Plan Nacional del Espacio under the projects ESP2001-4527-PE and ESP2001-4528-PE. PJA acknowledges financial support at the Instituto de Astrof\'{\i}sica de Andaluc\'{\i}a-CSIC 
by an I3P contract (I3P-PC2001-1) funded by the European Social Fund.

\begin{thebibliography}{}

\bibitem[Baglin et al. 2002]{baglin} Baglin A., Auvergne M., Barge P., Buey J.-T., Catala C., Michel E., Weiss W.~W., \& the COROT Team 2002, in Proceedings of the First Eddington Workshop on Stellar Structure and Habitable Planet Finding, ed. B. Battrick, Scientific editors: F. Favata, I. W. Roxburgh \& D. Galadi. ESA SP-485, 17

\bibitem[1992]{Baudrand} Baudrand, J., \& Bohm, T. 1992, A\&A, 259, 711

\bibitem[1997]{donati} Donati, J.~F., Semel, M., Carter, B.~D., Rees, D.~E., \& Cameron, A.C. 1997, MNRAS, 291, 658

\bibitem[1976]{Gron76} Gronbech, B., Olsen, E.~H., \& Str\"omgren, B. 1976, A\&AS, 26, 155 

\bibitem[Kurucz 1979]{kur} Kurucz R.L. 1979, ApJS, 40, 1

\bibitem[Neiner et al. 2004]{neiner} Neiner, C., Huber,t A.-M., \& Catala, C. 2004, ApJ, submitted

\bibitem[1993]{Olse93} Olsen, E.~H. 1993, A\&AS, 102, 89

\bibitem[1994a]{Olse94V} Olsen, E.~H. 1994a, A\&AS, 104, 429

\bibitem[1994b]{Olse94} Olsen, E.~H. 1994b A\&AS, 106, 257

\bibitem[2003]{Poretti} Poretti, E., Garrido, R., Amado, P.~J., Uytterhoeven, K., Handler, G., et al. 2003 A\&A 406, 203

\bibitem[2003]{Rainer03} Rainer, M. 2003, Laurea Thesis, Universit\'{a} di Milano (in Italian)
\bibitem[Rogers 1995]{rogers} Rogers, N.~Y. 1995, CoAst, 78

\bibitem[2003]{Tsymbal03} Tsymbal, V., Lyashko, D., \& Weiss, W.~W. 2003, in Processing Stellar Echelle Spectra, IAU Symp 210, editors. N. Piskunov, W.W.Weiss, D.F. Gray, Astr. Soc. Pacific.

\end{thebibliography}
\end{document}